# Structural, electronic, and optical properties of 6H-SiC layers synthesized by implantation of carbon ions into silicon

D.W. Boukhvalov[1,2*], D.A. Zatsepin[2,3], D.Yu. Biryukov[2],
Yu.V. Shchapova[2], N.V. Gavrilov[2], A.F. Zatsepin[2]

[1] *College of Science, Institute of Materials Physics and Chemistry, Nanjing Forestry University, Nanjing, P. R. China*
[2] *Institute of Physics and Technology, Ural Federal University, Ekaterinburg, Russia*
[3] *Institute of Metal Physics, Ural Branch of Russian Academy of Sciences, Ekaterinburg, Russia*

**Abstract**

Systematic studies of the gradual fabrication by means of carbon ion-implantation of high-quality 6H-SiC layers on silicon surfaces have been carried out. The fluence of carbon ions varied from $5\times10^{15}$ cm$^{-2}$ to $10^{17}$ cm$^{-2}$. Results of first-principle calculations, X-ray diffraction (XRD), and Raman spectroscopy demonstrate the amorphization of silicon substrate without any tendency to the segregation of carbon in the samples synthesized at low fluencies. The formation of a SiO$_2$-like structure at this stage was also detected. X-ray photoelectron spectroscopy (XPS), XRD, and Raman spectroscopy demonstrate that an increase in carbon content at $10^{17}$ cm$^{-2}$ fluence leads to the growth of 6H-SiC films on the surface of the amorphous silicon substrate. Atomic force microscopy (AFM) data obtained also demonstrates the decreasing of surface roughens after the formation of SiC film. XPS and Raman spectra suggest that excessive carbon content leaves the SiC matrix via the formation of an insignificant amount of partially oxidized carbon nanostructures. Optical measurements also support the claim of high-quality 6H-SiC film formation in the samples synthesized at $10^{17}$ cm$^{-2}$ fluence and demonstrate the absence of any detectable contribution of nanostructures formed from excessive carbon on the optical properties of the material under study.

E-mail: danil@njfu.edu.cn

# 1. Introduction

In the last decades, special attention has been paid to oxygen-free materials with a covalent type of interlayer interaction, a typical representative of which silicon carbide SiC is considered. Up to date, the existence of about 200 structural one-dimensional polymorphs – polytypes of a bulk precursor of silicon carbide has been confirmed, which is the base for applicative functionalization of SiC and allows to yield the broad set of physicochemical properties for the final functionalized material (see, e.g., Refs. [1–5]). Also, amorphous a-SiC polytypes are known [6]. Moreover, an essential feature of silicon carbide and structures based on SiC is the formation of properties that are significantly different from that of the bulk when the structural dimensions of the material are essentially reduced [7]. The latter opens up vast opportunities for creating composite materials with the same chemical formula but different functionalities. For this reason, the above features allow us to consider silicon carbide as an essential object of physical functionalization for various application areas and indicate the need for its onward study since SiC is the only material in which different crystal structures exhibit significantly different electronic properties and different electronic effective masses [8].

Different SiC polytypes have dissimilar stacking sequences of Si–C double layers. Hence, the band gap usually varies in the range from 2.4 to 3.3 eV, where nearly the medium value (~ 3.0 eV) is exhibited by 6H-SiC polytype with relatively good electron mobility (see, e.g., Ref. [9]). Formally, 6H-SiC polytype is one of the most easily synthesized if compared to others since the phenomenon of polytypism for silicon carbide negatively affects, in general, the synthesis of a single-phase final product. In this regard, both a reasonable selection of synthesis method and complex inspection by various experimental methods are highly likely required in order to control the single-phase structure of a final product. Such a complex approach is not a trivial scientific and technical task, even when synthesizing high-grade 6H-SiC polytype.

Therefore, the development of less sophisticated methods for the fabrication of surface SiC films is essential for the onward progress in this area of interest. Understanding the gradual formation process of the SiC surface layer is an important step toward developing more controlled techniques for modifying the surface of semiconductors. In the current paper, we are reporting the results of systematic experimental and theoretical studies of carbon-rich films on the silicon substrate after being implanted by carbon ions employing four different fluences.

## 2. Experimental and Theoretical Methods

In order to produce a carbon-doped silicon sample, the silicon wafer with bright and matte surfaces measuring 10×20 mm was used. The initial samples for implantation were polished wafers of silicon single crystals from Global Wafers Co., Ltd. (Product name: 6" Polishing Wafer; Product Method: CZ; Crystallographic orientation <100>). Prior to carbon implantation, this silicon wafer was etched with argon-ion for 10 minutes, employing 300 W high-frequency (13.56 MHz) discharge. The following pulsed carbon ion-implantation mode was used: pulse repetition rate was 6.25 Hz with 0.4 milliseconds of pulse duration, ion beam pulsed current density 1.9 mA/cm$^2$, ion energy 25 keV. The ion fluence varied from $5\times10^{15}$ to $1\times10^{17}$ cm$^{-2}$ by changing the duration of the implantation process; the maximum ion fluence of $1\times10^{17}$ cm$^{-2}$ was reached in 4080 seconds. Oil-free pumped vacuum in the vacuum chamber during ion-implantation was not worse than $2\times10^{-5}$ Torr. After setting the required value of ion fluence, the sample was kept in a vacuum chamber for 12 minutes prior to placing it into a humidity-protected transfer vessel. The surface topography after implantation was determined by employing Atomic Force Microscopy (AFM) using the NTEGRA Prima device (NT-MDT LLC, Russia).

Considering possible variations in the chemical composition and structure of the near-surface layer of a silicon wafer after carbon implantation, we measured the Raman spectra of the initial substrate and the sample under study. These spectra were obtained in backscattering geometry with the help of Horiba LabRam HR800 Evolution spectrometer equipped with an Olympus BX-FM confocal microscope employing 100x/NA = 0.9 objective and 50 μm confocal aperture. Excitation was performed using Ar-laser at λ=514.5 nm wavelength. Registration was carried out through a Cherny-Turner monochromator using diffraction grating with 1800 lines/mm and a multichannel-cooled CCD detector. The spectral resolution of the employed device under specified conditions was 0.6 см$^{-1}$, and spatial depth resolution d$_f$ (the size of the focal region) was 2 μm. The thickness of the analyzed layer, depending on d$_f$ values and the optical absorption coefficient of silicon at 514 nm wavelength (1/α ~ 1 μm), was a few μm, which significantly exceeded the penetration depth of C$^+$ ions (E = 25 keV) into silicon (~ 0.1 – 0.2 μm). Thus, the Raman spectra of the sample under study are the superposition of the spectra of the substrate and modified surface; the latter contribution was determined based on layer-by-layer Raman measurements with changing focusing z-depth.

We employed a Thermo Scientific K Alpha+ X-ray photoelectron spectrometer for taking X-Ray photoelectron spectra. The survey spectrum was recorded in a fast scan mode with Al $K\alpha$ excitation source, and 200 eV energy pass-window of 180° hemispherical energy analyzer with 0.05 at.% elements sensitivity. The vacuum in the analytical chamber of the employed spectrometer was $1.3 \times 10^{-6}$ Pa. An in-built XPS Database of spectrometers with cross-referencing to NIST XPS Standard Reference Database 20 (version 4.1) [10] and Handbook of Monochromatic XPS Spectra [11] was used in order to interpret the recorded survey spectrum.

Structural-phase analysis of the sample under study was performed employing XPertPro MPD diffractometer with Cu $K\alpha$ = 1.54 Å radiation. Optical reflection spectra were recorded using Perkin Elmer Lambda 35 spectrophotometer with an integrating sphere in the range of 190 – 1100 nm. The slit had 1 nm width, and the spectrum recording velocity was 270 nm/min. The wavelength setting error was ± 0.5 nm with confidence probability value P = 0.95. The limits of permissible absolute error of the employed spectrophotometer were DR = ± 0.5% when measuring the reflectance coefficient of the sample under study.

Scenarios for the initial stages of carbon embedding into oxidized silicon surfaces were studied by means of theoretical calculations using the DFT method. For this purpose a pseudopotential approach implemented in SIESTA code was used [13]. All calculations used Perdew-Burke-Ernzerhof generalized gradient approximation (GGA-PBE) with spin-polarization [14]. Relevant codes were employed in order to optimize final atomic positions completely. The ion cores were described during optimization by non-relativistic and norm-conserving pseudopotentials with cut-off silicon, carbon, oxygen, and hydrogen radii as 2.09, 1.47, 1.14, and 1.15 a.u., respectively [15]. Moreover, the wave functions were expanded with a double-ζ plus polarization basis of localized orbitals for all species (excluding hydrogen), while a double-ζ basis was applied for hydrogen. The k-point mesh within the Monkhorst-Pack scheme [16] is 6×6×2. Formation energies have been calculated employing the standard formula:

$$E_{form} = \{E(host + n\ guests) – [E(host) + nE(guest)]\}/n, \qquad (1)$$

where E(host) and E(host + n guests) are the total energy of the system before and after incorporation of n impurities. As E(guest), we used the total energy per atom in graphite with AB stacking. The negative sign of formation energy corresponds to exothermic processes, and the positive with endothermic.

## 3. Results and Discussions

### 3.1 X-ray diffraction analysis

XRD patterns shown in Fig. 1a demonstrate the presence of a dominating band located at 70 degrees, which corresponds to the reflection of x-rays from (400) silicon plane since the depth of carbon penetration is limited to ~ 50 nm. The diffraction pattern also exhibits bands at reflection angles of 33, 62, and 75 degrees, which are related to (101), (108), and (0012) planes of 6H-SiC structure [12]. These peaks indicate that synthesized SiC film has hexagonal structural modification.

The broad peak between 30 and 24 degrees observed in the samples synthesized at lower fluences can be discussed as the signature from layered carbon nanostructures [17] or nanosized silicon oxide [18]. The following simulations and measurements have been carried out to reveal the nature of this broad peak. Insignificant changes in width and the shift of the crystalline Si-related peak (about 0.1 degree) demonstrate the stability of atomic structure in subsurface area even for higher fluences.

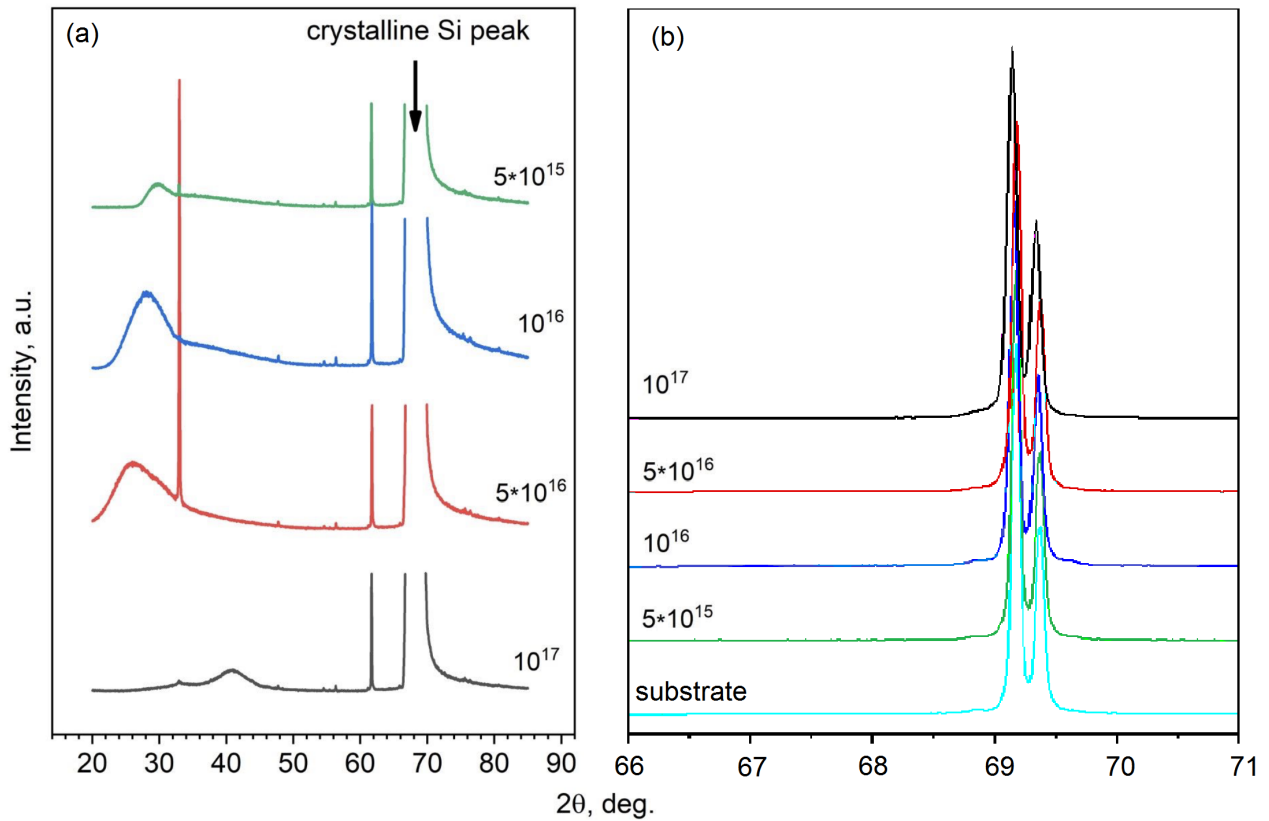

**Figure 1.** X-ray diffraction (XRD) patterns measured on Si:C samples synthesized at different fluences (a). XRD patterns for the same samples and non-treated substrates in the area corresponding with the c-Si peak are shown in panel (b).

## 3.2. DFT calculations

In order to inspect the possibility of carbon impurities segregation in the subsurface area of silicon matrices, we simulate atom-by-atom incorporation of carbon defects into the model slab corresponding to (111) surface of cubic silicon shown in Fig. 2a. To imitate the natural oxidation of silicon surface, we saturate all dangling bonds on the "top" part of the slab with hydroxyl groups. Hydrogen atoms saturated all dangling bonds from the 'bottom' side of the slab. The first step of our study is to inspect the possible pathway of single carbon atoms through interstitial voids of surface and subsurface layers. For this purpose, we located single interstitial carbon impurities in the center of silicon triangles on different levels and then performed optimization of atomic positions. Calculated atomic structures and formation energies are shown in Fig. 2b-d.

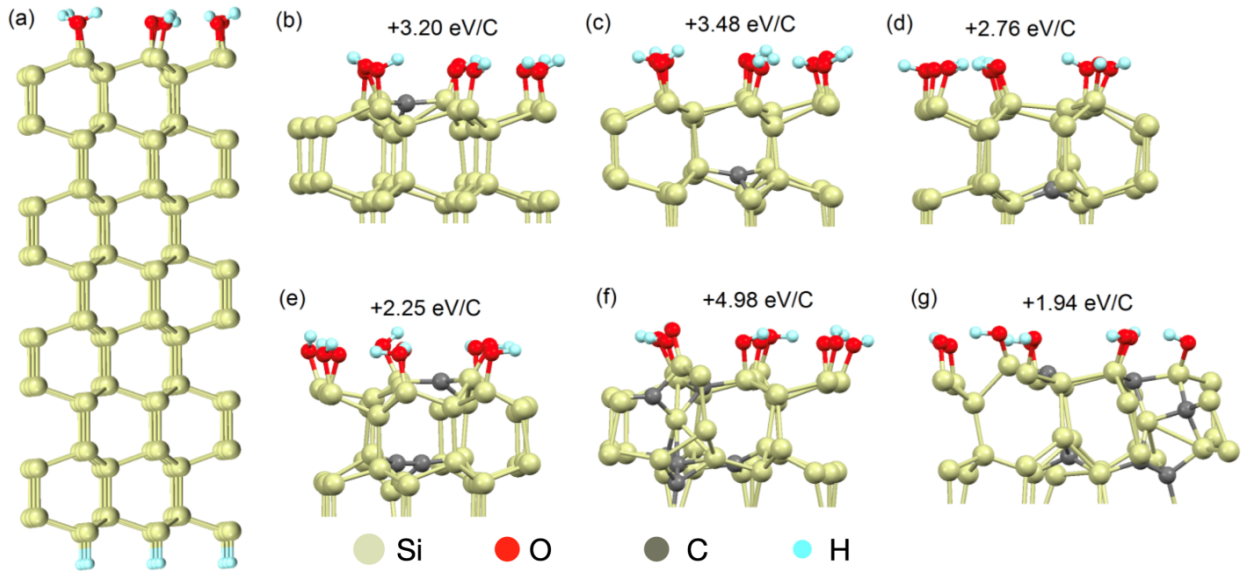

**Figure 2.** Optimized atomic structure of a whole supercell used for the simulations (a) and magnified near-surface area with inserted single carbon atoms in different positions (b-d), three carbon atoms in the surface and sub-surface area (e), and two different configurations of six carbon atoms in this area corresponding with the formation of multiple C-C bonds (f) and Si-C bonds (g).

Results of performed calculations demonstrate the favorability of single carbon impurity penetration into the subsurface area with the formation of three C–Si bonds with the nearest silicon atoms (see Fig. 2d). The next step of our calculations was to inspect the onward increase in the concentration of carbon impurities. For this purpose, we insert three carbon impurities into silicon's surface and subsurface areas. Results of calculations (see Fig. 2e) demonstrate the formation of C–C bonds and the favorability of this configuration relative to the most energetically favorable structure shown in Fig. 2d.

At the next step of our simulation, two additional carbon impurities were added to the structure shown in Fig. 2e. Two different starting configurations of additional carbon impurities pair were considered. The results of calculations shown in Fig. 2f-g demonstrate two distinct scenarios. In the first case shown in Fig. 2f, we have observed the formation of multiple carbon-carbon bonds. In the second case (see Fig. 2g), the formation of Si-C bonds is shown. Since the second scenario corresponds to significantly lower (about 3 eV/C) formation energy, we can propose this route as more probable. Thus, theoretical simulation of step-by-step incorporation of interstitial carbon impurities into silicon's oxidized surface demonstrates the favorability of forming Si-C-Si-C patterns, which can be proposed as seeds of further silicon carbide surface phase. Contrarily, the segregation of carbon atoms with the formation of carbon clusters in subsurface areas is a significantly less favorable process.

### 3.3. Raman spectra

Raman spectra of silicon substrate and carbon implanted samples at lower fluencies demonstrate relatively broad (FWHM = 70~100 cm$^{-1}$) maxima at ~150, ~300, ~430, ~480 cm$^{-1}$, and narrow (FWHM=6-7 cm$^{-1}$) peak at 520 cm$^{-1}$ (see Fig. 3a). The shape and location of broad bands correspond to TA, LA, LO, and TO vibrational modes of amorphous silicon a-Si, whereas 520 cm$^{-1}$ maximum corresponds to the vibrational mode of crystalline silicon c-Si [19,20]. Layer-by-layer Raman measurements of silicon substrates showed that they are partially amorphous throughout the entire thickness of the analyzed layer. If the fluence is increased, then the narrow 520 cm$^{-1}$ peak of c-Si naturally decreases. The detected amorphization of silicon substrate [21-23] is in agreement with the formation of large disordered areas at the initial stages of carbon incorporation, as predicted by theory (see Fig. 2g). Increasing intensity in the area 750-1000 cm$^{-1}$ can correspond to the formation of SiO$_2$-like structures. Similar changes in Raman spectra have been observed in Si-implanted SiO$_2$ [24] and SiO$_2$@NiO heterostructures [25]. Note that the formation of these SiO$_2$-like structures in the studied samples was proposed based on XRD results.

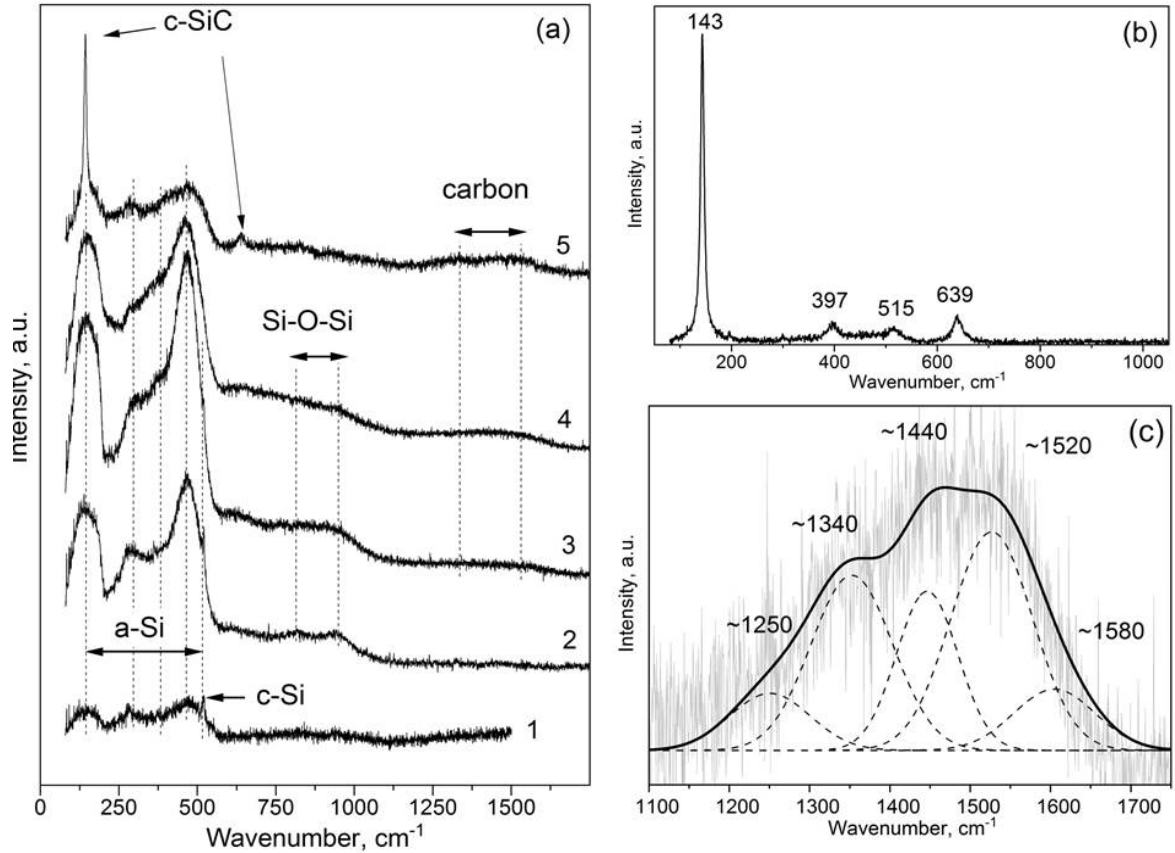

**Figure 3.** Raman spectra of silicon substrate and carbon implanted samples (a) corresponding to 1 – non-implanted substrate, 2 – $1\times10^{16}$, 3 – $5\times10^{16}$, 4 – $1\times10^{17}$ (surface layer), 5 – $1\times10^{17}$ (subsurface layer). Panel (b) shows Raman spectra of $10^{17}$ cm$^{-2}$ sample after subtraction of silicon substrate contribution. Raman signal from carbon-carbon bonds area in samples which were modified by $1\times10^{17}$ fluence (c).

The formation of a new phase clearly seen in Raman spectra for the samples fabricated employing $10^{17}$ cm$^{-2}$ fluence (see Fig. 3b). Formation of this new surface phase corresponds to the appearance of relatively narrow and intensive Raman peak located at 143 cm$^{-1}$ (FWHM=8.5 cm$^{-1}$) and accompanying low-intensity peaks at 397, 515 cm$^{-1}$ (FWHM= 38 cm$^{-1}$) and 639 cm$^{-1}$ (FWHM=22 cm$^{-1}$) (see Fig. 3b). Taking into account the small width of these maxima, we can assume that this feature is associated with appropriate crystalline phases. The peak located at 143 cm$^{-1}$ corresponds to $E_2$(TA) vibrations in the hexagonal wurtzite-like structure of crystalline 6H-SiC [26,27]. A low-intensity maximum at 515 cm$^{-1}$ can be assumed as the manifestation of $A_1$(LA) vibrations of 6H-SiC and TA vibrations of crystalline silicon [17]. However, the latter is unlikely due to the amorphization of the silicon substrate in this sample noted above. The appearance of crystalline SiC phase in the surface layer of silicon after carbon implantation is consistent with the data reported in Refs. [28-33]. The broadened peak at 639 cm$^{-1}$ can be attributed to Si–C vibrations in the distorted structure of silicon carbide [28]. Despite clear manifestation of 143 cm$^{-1}$ peak, some additional crystal modes $E_2$(PO) at 767.5 and 788 cm$^{-1}$, $E_1$(TO) at 796 cm$^{-1}$, $A_1$(AO) at 888.5 cm$^{-1}$, $A_1$(LO) at 966.5 cm$^{-1}$ discussed in Refs. [26,32] were not detected.

Raman spectra of carbon implanted samples in 1000~1700 cm$^{-1}$ range also display a shape that is typical for the vibration spectrum of carbon in the *sp$^2$* and *sp$^3$* hybridized states (Fig. 3c). These low-intensity spectra were recorded only for the samples implanted with higher carbon content (fluencies $5\times10^{16}$ cm$^{-2}$ and $10^{17}$ cm$^{-2}$). At the same time, the signal in this region for the samples implanted with lower fluencies ($5\times10^{15}$ cm$^{-2}$ and $10^{16}$ cm$^{-2}$) was absent. The spectra exhibit several broad bands, which are usually associated with C–C vibrations of graphite (G-band located in the range of 1580–1600 cm$^{-1}$), disordered *sp$^2$*- and *sp$^3$*-carbon (D-band located at ~1350 cm$^{-1}$), and other defect states of carbon matter (see Ref. [33]). The results obtained show that, at least, for samples implanted with the highest fluence ($10^{17}$ cm$^{-2}$), carbon particles (clusters) were fabricated. The observed lack of ordering of carbon clusters is in agreement with the absence of typical for ordered carbon structures' distinct peak at about 26 degrees in XRD patterns of the sample fabricated at $10^{17}$ cm$^{-2}$ fluence. These results are in agreement with the theoretically predicted lack of carbon impurities clustering in the subsurface area of silicon at the initial stages of carbon embedding (see section 3.2). Thus, the broad peak between 30 and 24 degrees detected in XRD patterns for the samples with low carbon content (see Fig. 1b) can be interpreted as a peak associated with the formation of silica-like structures, which disappear after the formation of crystalline SiC surface film.

### 3.4. Surface topography

Our next research stage is to determine the quality of the resulting surface films using Atomic Force Microscopy (AFM). It was found that the topography of initial (before ion-implantation) and implanted surfaces are different for smooth and matte silicon surfaces. The surface of samples after ion-implantation acquires a characteristic relief in the form of "craters" and "peaks", which have a diameter of ~ 100 nm and a depth or height of about 15~20 nm in samples implanted with $5\times10^{15}$ to $1\times10^{16}$ cm$^{-2}$ fluencies (see Fig. 4a,b). Further increase of carbon content upon implantation with $1\times10^{17}$ fluence leads to the formation of visible and more uniform surface films shown in Fig. 4c. The formation of crystalline SiC film leads to an observable improvement in the quality of the final surface. The observed surface morphology is consistent with theoretical predictions and Raman spectra discussed in previous sections.

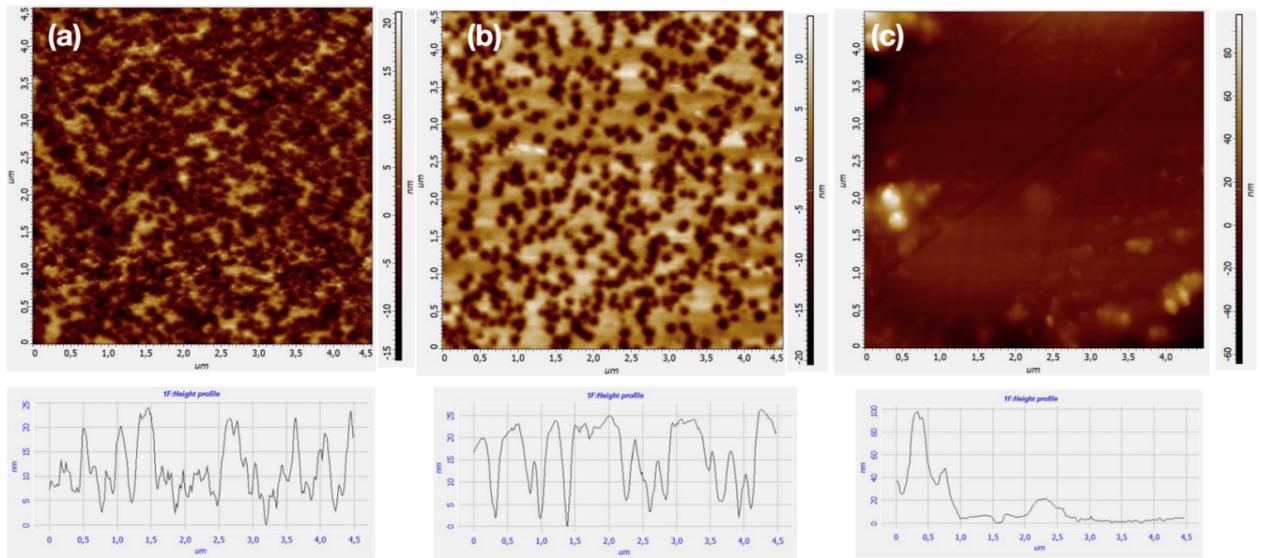

**Figure 4.** 2D AFM images of surface and depth profiles for samples synthesized using fluences $5\times10^{15}$ cm$^{-2}$ (a), $1\times10^{16}$ cm$^{-2}$ (b), and $1\times10^{17}$ cm$^{-2}$ (c) after correcting initial relief by subtracting second-order surface.

*3.5. XPS core-level spectroscopy*

Since oxygen-free samples can be accidentally oxidized in air, we performed a chemical elemental analysis of the synthesized Si:C sample using photoelectron spectroscopy. XPS elemental analysis allows us to conclude that the very weak oxidation of Si:C sample surface occurred (see weak XPS O 1s signal in Fig.5a). Measured concentration of this oxygen is not more than 1.36 at. %. At the same time, no other contaminators were found within the elemental sensitivity of the employed XPS spectrometer, what is indicating a relatively high-quality grade of the synthesized sample.

Since the formation of high-quality SiC films was detected only for the samples fabricated at the highest fluence of carbon ions (i.e., $10^{17}$ cm$^{-2}$), only these samples were the objects for XPS measurements. The main maximum in Si *2*p spectrum shown in Fig. 5b reflects the dominating Si–C bonding, whereas slightly asymmetrical high-energy band-tail originates due to Si–O bond contribution [10,35]. XPS identification obtained does not contradict the XPS survey spectrum (Fig.5a), where weak and irregular oxidation with 1.36 at.% of absorbed oxygen was established.

Figure 5b shows the XPS O *1*s core-level spectrum of the Si:C sample and a-SiO$_2$ XPS External Reference Standard. It is known from NIST XPS Standard Reference Database 20 (version 4.1) [10] and Thermo Scientific XPS Knowledge Base [35] that silicon-oxygen compounds have 531.9 eV binding energy position for O *1*s core-level (what can be seen in Fig. 5c), whereas slightly oxidized carbonates of silicon exhibit their weak O *1*s spectrum in the form of relatively broad shape at 531.9–532 eV.

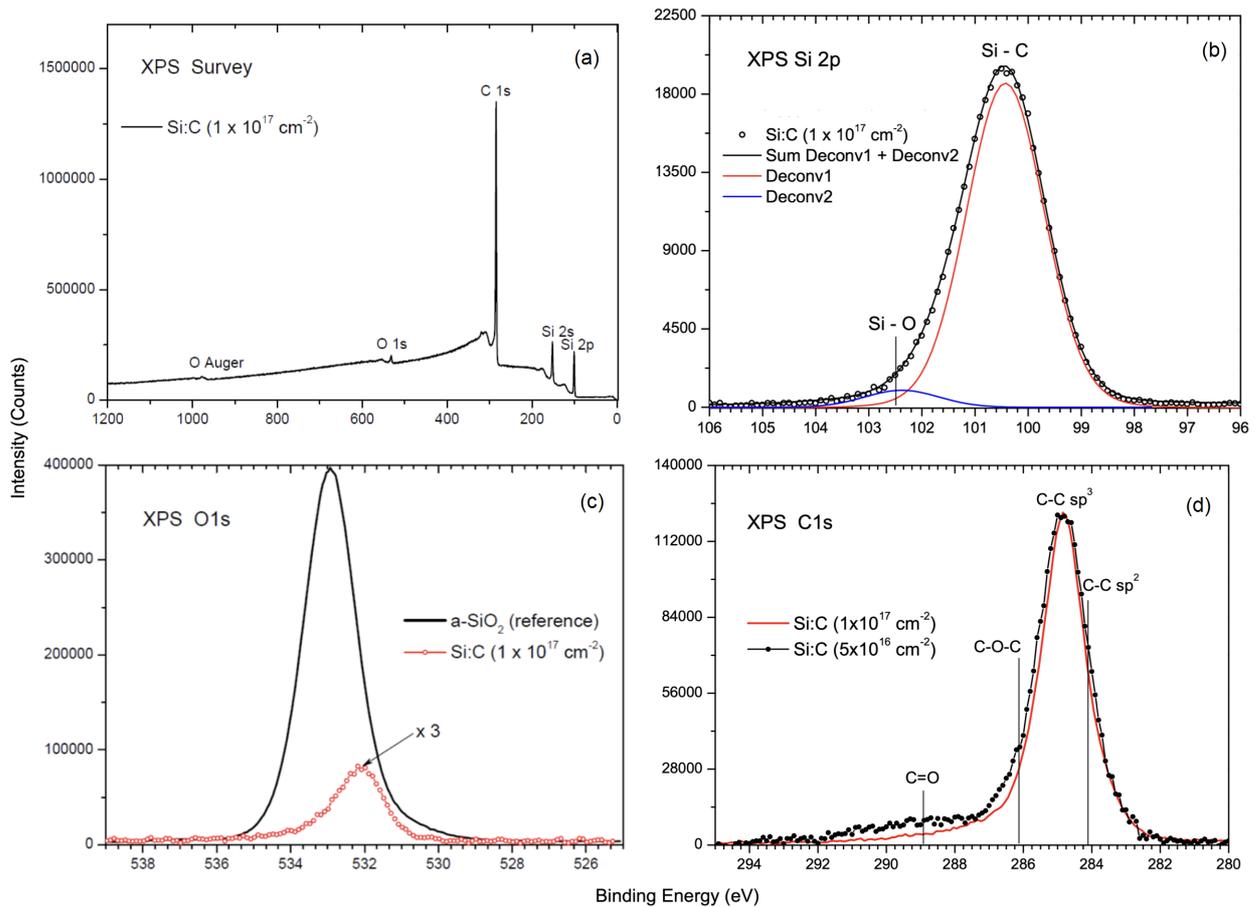

**Figure 5.** X-ray photoelectron survey spectrum of Si:C sample doped by $1\times10^{17}$ cm$^{-2}$ ion fluence (a), XPS Si *2*p (b), O *1*s (c), and C *1*s (d) core levels of Si:C sample fabricated using $10^{17}$ cm$^{-2}$ carbon ion fluence.

Our XPS data shown in Figure 5c well coincides with that reported by NIST XPS Standard Reference Database 20 (version 4.1) [10] and Thermo Scientific XPS Knowledge Base [34], thus allowing us to conclude that formally only carbon had formed bonding with oxygen and not the silicon. NIST XPS Standard Reference Database 20 (version 4.1) [10] also reports that C–O and C=O bonds contribute to O *1*s core-level at 531.9–532 eV, which also supports our viewpoint about partial irregular oxidation of carbon in the studied sample. The described absence of valuable oxidation of silicon revealed by Si *2*p and O *1*s spectra is in agreement with the absence of $SiO_2$-related peak in XRD patterns (Fig. 1b) measured for the samples fabricated at the highest fluence.

Figure 5d displays XPS C *1*s core-level spectrum of the Si:C sample doped by $10^{17}$ cm$^{-2}$ ion fluence. This figure shows that C *1*s spectrum exhibits a symmetrical shape with mixed *sp²-sp³* hybridization where *sp³* has a majority [10,36]. The same is reporting XPS spectrometer in-built XPS Database. At the same time, there are contributions from carbon-oxygen irregular clusters with C-O-C and C=O bonds, which arise at ~286 eV and ~289 eV [36], indicating carbon partial oxidation. In order to inspect the appearance of carbon structures in the sample with lower carbon content, C *1*s spectrum was recorded for the sample implanted with $5\times10^{16}$ cm$^{-2}$ ion fluence. This measured spectrum coincides well with that of the sample implanted with the highest carbon content. Thus, we can conclude that the broad peak between 30 and 24 degrees in XRD patterns of the samples implanted with low carbon content (see Fig. 1) was definitely caused by the formation of $SiO_2$-like structures. The disappearance of these structures in the sample with the highest carbon content can be explained by removing oxygen atoms from the solid phases by forming CO and $CO_2$ molecules. Note that the contribution from C-O-C and C=O bonds is increased with increasing carbon content; we can propose that these oxidized carbon atoms belong to some carbon nanostructures discussed in section 3.3. Thus, in addition to the formation of surface SiC films, implanted carbon atoms play a role in reducing oxygen content in these films.

*3.6. Optical properties*

Samples fabricated employing $10^{16}$ cm$^{-2}$ and $10^{17}$ cm$^{-2}$ carbon-ions fluences have been selected to inspect the quality effect on optical properties and possible formation of residual $SiO_2$-like and carbon-based structures. Dielectric SiC films have a much larger optical transmission gap than semiconductor silicon substrates. In this regard, the reflectance spectra of samples were measured. In this case, spectral dependences of SiC film absorption coefficient can be obtained by analyzing experimental data of reflection coefficient *R(hν)* in accordance with the model proposed by Kubelka-Munk [37]:

$$F(h) = \frac{(1 - R(h\nu))^2}{2R(h\nu)} \qquad (2)$$

where *F(hv)* denotes the Kubelka-Munk function, which is directly proportional to the absorption coefficient. Absorption spectra obtained from reflectance spectra using equation (2) are shown in Fig. 6a.

The resulting absorption spectra were analyzed in the region of the density of states for band tails using the Urbach rule [38]:

$$F(h\nu) = F_0 \cdot exp\left(\frac{h\nu - E_0}{E_U}\right) \qquad (3)$$

where $F_0$ and $E_0$ are the coordinates of the intersection point of the so-called "crystal-like" Urbach rule, and EU is Urbach energy, characterizing the measure of the general structural disorder of the system under study (static and dynamic disordering) [39]. Urbach absorption edge is formed due to optical transitions between localized electronic states, which are caused by corresponding distortions of atomic structure. In our case, $E_U$ parameter is a measure of the overall disorder of the material. Error-values were determined with the help of data taken from experimental dependencies approximation employing OriginPro 2018 SR1 v9.5.1.195 software.

The crystal-like "fan-like" behavior of absorption spectra for the SiC matrix indicates its crystalline structure, which is well consistent with XRD data. A similar nature of the Urbach rule was previously observed for transparent dielectrics after irradiation with various doses of accelerated ions [39]. From Figure 6a, one can see that the coordinate of the crossing-point in spectra is determined by the values $F_0$ = 0.932 arb. units and $E_0$ = 3.850 eV. At the same time, Urbach energy $E_U$ decreases from 741.7±0.5 to 571.7±0.5 meV (Fig. 7a) in films with an increase of carbon implantation fluence from $10^{16}$ up to $10^{17}$ cm$^{-2}$. This behavior indicates an increase in atomic ordering in the film structure. Such a fact can be explained by the improvement of stoichiometry under the influence of higher doses of carbon ions and the correction of predominantly static radiation damage (radiation annealing) in the lattice of the final 6H-SiC structure.

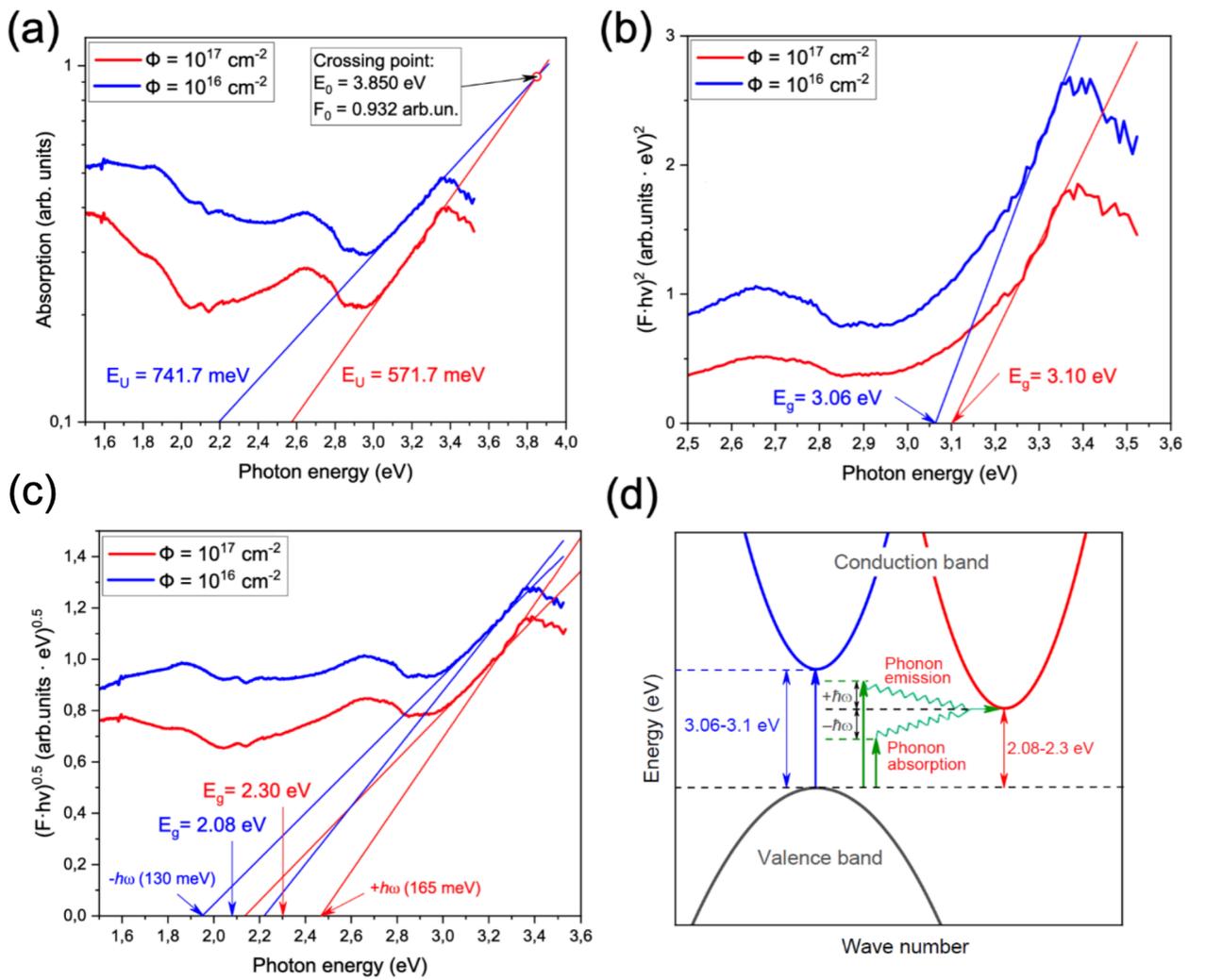

**Figure 6.** Optical absorption spectra of SiC film on Si substrate implanted with carbon ions using $10^{16}$ cm$^{-2}$ and $10^{17}$ cm$^{-2}$ fluences. Thin lines depict approximations of an exponential portion of the spectrum employing the Urbach rule (a). The same spectra were plotted in Tauc coordinates for direct allowed transitions (b) and indirect transitions (c). Schematic representation of direct (blue arrows) and indirect (red arrows) transitions in 6H-SiC film (d). The green wave-shaped arrows show the processes of absorption and emission of phonons during indirect transitions.

In order to determine energy gap values, we analyzed spectral dependences of absorption determined using the Kubelka-Munk approach in Tauc coordinates [40,41] for direct and indirect optical transitions. The Tauc equation is represented by expression (4):

$$(F \bullet h\nu)^n = A \bullet (h\nu - E_g) \qquad (4)$$

where $A$ is a constant, F is the Kubelka-Munk function, h$\nu$ denotes photon energy, $E_g$ is the value of the optical transparency gap, and $n$ is a parameter depending on the type of transition.

Within the framework of direct allowed transitions model (i.e., n = 2), the width of optical transparency gap was determined by extrapolating linear portion of $(F \cdot h\nu)^2$ function to the intersection with abscissa axis, as shown in Fig. 6b. The obtained values of direct optical gap in the carbon-implanted layers of Si substrate are 3.06 ± 0.02 eV and 3.10 ± 0.02 eV for $10^{16}$ and $10^{17}$ cm$^{-2}$ fluences, respectively. The observed increase in the energy of interband optical transitions is associated with a slight decrease in the length of band tails during atomic structure ordering of the final film. As mentioned above, the fact of the ion-induced ordering of the structure is most noticeably recorded by the decrease in Urbach energy (Fig. 6a). A surface disordered phase with disrupted stoichiometry is formed at the initial stages of the ion irradiation process at low carbon fluences (i.e., $10^{16}$ cm$^{-2}$). But at $10^{17}$ cm$^{-2}$ fluences the structural and chemical ordering occurs, which is accompanied by the shortening of band tails and an increase in the energy gap both for direct and indirect optical transitions. The effect of increasing optical gap correlates with a decrease in the Urbach energy, which characterizes the degree of general (static and dynamic) structural disorder in the system (Fig. 6a).

The presence of long band tails can facilitate not only direct but also indirect optical transitions. Within the framework of indirect transitions model (n = 0.5), two linear sections of $(F \cdot h\nu)^{0.5}$ function were analyzed, corresponding to $\pm \hbar \omega$ energy of phonons participating in the process of interband photothermal excitation of the electronic subsystem of film (Fig. 6c). The optical gap width for indirect transitions was 2.08±0.02 and 2.30±0.02 eV for $10^{16}$ and $10^{17}$ cm$^{-2}$ fluences, respectively. Moreover, the energy of phonons participating in indirect optical transitions is also increased in the range of 130 ± 5 to 165 ± 5 meV with an increase in ion fluence. The obtained values of $\hbar \omega$ are close to the theoretical values of *LO*-phonon energy in the perfect 6H-SiC structure [42]. A schematic representation of direct and indirect transitions with absorption and emission of *LO*-phonons in 6H-SiC film is shown in Fig. 6d. All results of analyzed absorption spectra of 6H-SiC films are summarized in Table 1.

**Table 1.** Energy, optical, and phonon parameters sensitive to the structure of 6H-SiC films

| $\Phi$, cm$^{-2}$ | $E_U$, meV | direct transitions (n = 2) | indirect transitions (n = 0.5) | |
|---|---|---|---|---|
| | | $E_g$, eV | $E_g$, eV | $\hbar \omega$, meV |
| $10^{16}$ | 741.7±0.5 | 3.06±0.02 | 2.08±0.02 | 130±5 |
| $10^{17}$ | 571.7±0.5 | 3.10±0.02 | 2.30±0.02 | 165±5 |

The presented above table clearly shows that such parameters as Urbach energy, value of indirect optical gap and energy of *LO*-phonons involved in interband transitions are the most sensitive to the dose of carbon-ion irradiation and film structural state. The energy of direct optical gap at different fluences varies to a much lesser extent. These features of optical properties of the film under study can be interpreted as a consequence of the influence of the dominating role of static atomic disorder with a partial contribution from crystal lattice dynamics. At the same time, pure electronic transitions without the participation of the phonon subsystem depend, to a lesser extent, on the conditions for the film structure formation.

**Conclusions**

Systematic studies of silicon surfaces doped by carbon ions employing different fluences have been carried out. Results of simulations and characterizations demonstrate that amorphization of silicon and the formation of $SiO_2$-like surface structures occurs at low fluences of implantation. This process corresponds to increasing visible roughness of the surface observed by AFM. The onward increase of fluence up to $10^{17}$ cm$^{-2}$ led to the formation of a high-quality 6H-SiC surface layer with the disappearance of $SiO_2$-like secondary phases. Described the gradual transition from crystalline through the amorphous surface to the formation of the ordered surface film can be used for the guided modification of semiconductor' surfaces. On the other hand, the formation of an amorphous (a-Si) layer observed at lower fluences is important for the development of surface doped-a-Si/c-Si interfaces.

A high degree of crystallinity of this 6H-SiC film has been confirmed by XRD and Raman spectra measurements. XPS also demonstrates the formation of high-quality weakly-oxidized SiC surface film. XPS and Raman spectra reveal that excessive carbon forms a tiny number of oxidized disordered carbon-based nanostructures, which does not affect the structural properties of 6H-SiC surface films. On the other hand, excessive carbon can be described as an efficient oxygen harvester. This approach can also be used for the purification of surface areas of semiconductors. Optical measurements demonstrate the sensitivity of band energy structure and states of electronic and phonon subsystems to the conditions of 6H-SiC layers formation and atomic ordering-disordering processes.

The systematic description of the gradual process of high-quality 6H-SiC film formation on the surface of silicon bombarded by carbon ions can be useful in guiding the design of new specialized optoelectronic materials and functional instruments.

**Acknowledgments**

The study was supported by the Ministry of Science and Higher Education of the Russian Federation (Ural Federal University Program of Development within the Priority-2030 Program, project 4.38). The samples of SiC films were synthesized at the Institute of Electrophysics, Ural Branch of the Russian Academy of Sciences. The equipment of Ural Center ''Modern Nanotechnologies'' of Ural Federal University (Reg. 2968) was used for samples characterization. Raman spectra were taken at "Geoanalitik" shared research facilities of FSBIS Zavaritsky Institute of Geology and Geochemistry of the Ural Branch of the Russian Academy.

## References

[1] A.L. Falk, P.V. Klimov, V. Ivády, K. Szász, D.J. Christle, W.F. Koehl, Á. Gali, D.D. Awschalom. Optical Polarization of Nuclear Spins in Silicon Carbide, Phys. Rev. Lett. 114 (2015) 247603. http://dx.doi.org/10.1103/PhysRevLett.114.247603

[2] B. Peng, Y. Zhang, Y. Wang, H. Guo, L. Yuan, R. Jia. Ferromagnetism Observed in Silicon-Carbide-Derived Carbon, Phys. Rev. B 97 (2018) 054401. https://doi.org/10.1103/PhysRevB.97.054401

[3] R. Nagy, M. Widmann, M. Niethammer, D.B.R. Dasari, I. Gerhardt, Ö.O. Soykal, M. Radulaski, T. Ohshima, J. Vučković, N.T. Son, I. G. Ivanov, S.E. Economou, C. Bonato, S.-Y. Lee, J. Wrachtrup. Quantum Properties of Dichroic Silicon Vacancies in Silicon Carbide, Phys. Rev. Applied 9 (2018) 034022. https://doi.org/10.1103/PhysRevApplied.9.034022

[4] N. Alaal, V. Loganathan, N. Medhekar, A. Shukla. From Half-Metal to Semiconductor: Electron-Correlation Effects in Zigzag SiC Nanoribbons From First Principles, Phys. Rev. Applied 7 (2017) 064009. https://doi.org/10.1103/PhysRevApplied.7.064009

[5] F.Z. Ramadan, H. Ouarrad, L. B. Drissi. Tuning Optoelectronic Properties of the Graphene-Based Quantum Dots $C_{16-x}Si_xH_{10}$ Family, J. Phys. Chem. A 122 (2018) 5016. https://doi.org/10.1021/acs.jpca.8b02704

[6] V.I. Ivashchenko, V.I. Shevchenko. Atomic and Electronic Structures of a-SiC, Semicond. Phys. Quant. Electronics & Optoelectronics 1 (2002) 16.

[7] A.V. Kalashnikov, A.V. Tuchin, L.A. Bityutskaja. Electronic Structure of Multilayer Allotropes of 2D Silicon Carbide, Letters on Materials 9 (2019) 173. https://doi.org/10.22226/2410-3535-2019-2-173-178

[8] F. Bechstedt, P. Käckell, A. Zywietz, K. Karch, B. Adolph, K. Tenelsen, J. Furthmüller. Polytypism and Properties of Silicon Carbide, Phys. Stat. Sol. B 202 (1997) 35. https://doi.org/10.1002/1521-3951%28199707%29202%3A1%3C35%3A%3AAID-PSSB35%3E3.0.CO%3B2-8

[9] M. Matos. Electronic Structure of Several Polytypes of SiC: A Study of Band Dispersion From A Semi-empirical Approach, Physica B 324 (2002) 15. https://doi.org/10.1016/S0921-4526(02)01203-6

[10] A.V. Naumkin, A. Kraut-Vass, S.W. Gaarenstroom, C.J. Powell. NIST XPS Standard Reference Database 20, version 4.1, © 2012, accessed 2023-09-23, https://doi.org/10.18434/T4T88K, https://srdata.nist.gov

[11] B.V. Crist. Handbook of Monochromatic XPS Spectra: The Elements of Native Oxides, Ed.: John Wiley & Sons LTD, NY, USA, ISBN: 978-0-471-49265-8 (2000) 548 p.

[12] A.V. Semenov, V.M. Puzikov, M.V. Dobrotvorskaya, A.G. Fedorov, A.V. Lopin. Nanocrystalline SiC Films Prepared by Direct Deposition of Carbon and Silicon Ions, Thin Solid Films 516 (2008) 2899. https://doi.org/10.1016/J.TSF.2007.05.059

[13] J.M. Soler, E. Artacho, J.D. Gale, A. Garsia, J. Junquera, P. Orejon, D. Sanchez-Portal. The SIESTA method for ab-initio order-N materials simulation, J. Phys.: Condens. Matter 14 (2002) 2745. https://doi.org/10.1088/0953-8984/14/11/302

[14] J.P. Perdew, K. Burke, M. Ernzerhof. Generalized gradient approximation made simple, Phys. Rev. Lett. 77 (1996) 3865. https://doi.org/10.1103/PhysRevLett.77.3865

[15] N. Troullier, J.L. Martins. Efficient pseudopotentials for plane-wave calculations, Phys. Rev. B 43 (1991) 1993. https://doi.org/10.1103/PhysRevB.43.1993

[16] H.J. Monkhorst, J.D. Pack. Special points for Brillouin-zone integrations. Phys. Rev. B 13 (1976) 5188. https://doi.org/10.1103/PhysRevB.13.5188

[17] Y. Yang, J. Hou, D. Huo, X. Wang, J. Li, G. Xu, M. Bian, Q. He, C. Hou, M. Yang, Green emitting carbon dots for sensitive fluorometric determination of cartap based on its aggregation effect on gold nanoparticles. Microchim. Acta 186 (2019) 259. https://link.springer.com/article/10.1007%2Fs00604-019-3361-5

[18] Y. Liang, J. Ouyang, H. Wang, W. Wang, P. Chui, K. Sun, Synthesis and characterization of core-shell structured $SiO_2@YVO_4:Yb^{3+},Er^{3+}$ microspheres, Appl. Surf. Sci. 258 (2012) 3689-3694. http://dx.doi.org/10.1016/j.apsusc.2011.12.006

[19] A. Zwick, R. Carles, Multiple-order Raman scattering in crystalline and amorphous silicon, Phys. Rev. B. 48 (1993) 6024. https://doi.org/10.1103/PhysRevB.48.6024

[20] Z. Li, W. Li, Y. Jiang, H. Cai, Y. Gong, J. He, Raman characterization of the structural evolution in amorphous and partially nanocrystalline hydrogenated silicon thin films prepared by PECVD, J. Raman Spec. 42 (2011) 415. https://doi.org/10.1002/jrs.2711

[21] D. Beeman, R. Tsu, M.F.Thorpe. Structural information from the Raman spectrum of amorphous silicon, Phys. Rev. B 32 (1985) 874. https://doi.org/10.1103/physrevb.32.874

[22] R.L.C. Vink, G.T. Barkema, W.F. van der Weg. Raman spectra and structure of amorphous Si, Phys. Rev. B 63 (2001) 1152010. https://doi.org/10.1103/PhysRevB.63.115210

[23] P. Yogi, M. Tanwar, S. K. Saxena, S. Mishra, D.K. Pathak, A. Chaudhary, P.R. Sagdeo R. Kumar. Quantifying the Short-Range Order in Amorphous Silicon by Raman Scattering, Analyt. Chem. 90 (2018) 8123. https://doi.org/10.1021/acs.analchem.8b01352

[24] S. B. Vishwakarma, S.K. Dubey, R.L. Dubey, Structural and Optical Investigations of $SiO_2$ Layers Implanted with 100 keV Silicon Negative Ions, JETIR 6 (2019) 148-154.

[25] A. Prasath A.S. Sharma, P.Elumalai, Nanostructured $SiO_2@NiO$ heterostructure derived from laboratory glass waste as anode material for lithium-ion battery, Ionics 25 (2019) 1015–1023. https://link.springer.com/article/10.1007/s11581-019-02879-9

[26] J.C. Burton, L. Sun, M. Pophristic, S.J. Lukacs, F.H. Long, Z.C. Feng, I.T. Ferguson Spatial characterization of doped SiC wafers by Raman spectroscopy, J. Appl. Phys. 84 (1998) 6268. https://doi.org/10.1063/1.368947

[27] G. Chikvaidze N. Mironova-Ulmane, A. Plaude, O. Sergeev, Investigation of silicon carbide polytypes by Raman spectroscopy, Latv. J. Phys. Tech. Sci. 3 (2014) 51. https://doi.org/10.2478/lpts-2014-0019

[28] J.A. Borders, S.T. Picraux, W. Beezhold. Formation of SiC in silicon by ion implantation, Appl. Phys. Letters, 18 (1971) 509. https://doi.org/10.1063/1.1653516

[29] D.J. Brink, J. Camassel, J. B. Malherbe. Formation of a surface SiC layer by carbon-ion implantation into silicon, Thin Solid Films, 449 (2004) 73. https://doi.org/10.1016/j.tsf.2003.10.018

[30] P.R. Poudel, P.P. Poudel, B.P. Sharma, J.Y. Hwang., M. Bouanani, B. Rout, F.D. McDaniel, Synthesis of buried layers of β-SiC in Si by multiple energy carbon ion implantations and post thermal annealing, Thin Solid Films 524 (2012) 35. https://doi.org/10.1016/j.tsf.2012.09.06

[31] J. Khamsuwan, S. Intarasiri, K. Kirkby, C. Jeynes, P.K. Chu, T. Kamwanna, L. D. Yu. High-energy heavy ion beam annealing effect on ion beam synthesis of silicon carbide, Surf. Coatings Technol. 206 (2011) 770. https://doi.org/10.1016/j.surfcoat.2011.04.058

[32] N. Chaâbane, A. Debelle, G. Sattonnay, P. Trocellier, Y. Serruys, L. Thomé, Y. Zhang, W.J. Weber, C. Meis, L. Gosmain, A. Boulle. Investigation of irradiation effects induced by self-ion in 6H-SiC combining RBS/C, Raman

and XRD, Nuclear Instr. Meth. Res. Sect. B: Beam Interact. Mater. Atoms. 286 (2012) 108. https://doi.org/10.1016/j.nimb.2011.11.018

[33] V.V. Artamonov, M.Y. Valakh, N.I. Klyui, V.P. Melnik, A.B. Romanyuk, B.N. Romanyuk, V.A. Yuhimchuk. Effect of oxygen on ion-beam induced synthesis of SiC in silicon, Nuclear Instruments and Methods in Physics Research Section B: Beam Interactions with Materials and Atoms, 147 (1999) 256. https://doi.org/10.1016/s0168-583x(98)00607-7

[34] A. Ferrari, J. Robertson. Interpretation of Raman spectra of disordered and amorphous carbon, Phys. Rev. B. 61 (2000) 14095. https://doi.org/10.1103/PhysRevB.61.14095

[35] Thermo Scientific XPS: Knowledge Base, © 2013–2021, accessed 2023-09-22, https://www.thermofisher.com/ru/ru/home/materials-science/learning-center/periodic-table/non-metal/oxygen.html

[36] Thermo Scientific XPS: Knowledge Base, © 2013–2021, accessed 2023-09-22, https://www.thermofisher.com/ru/ru/home/materials-science/learning-center/periodic-table/non-metal/carbon.html

[37] P. Kubelka, F. Munk, Ein Beitrag zur Optik der Farbanstriche. J. Tech. Phys. 12 (1931) 593–601. See also English translation by S. Westin (An article on optics of paint layers, http://www.graphics.cornell.edu/~westin/pubs/kubelka.pdf)

[38] F. Urbach, The long-wavelength edge of photographic sensitivity and of the electronic Absorption of Solids. Phys. Rev. 92 (1953) 1324. http://doi.org/10.1103/PhysRev.92.1324

[39] A.F. Zatsepin, D.Yu. Biryukov, D.A. Zatsepin, T.V. Shtang, N.V. Gavrilov. Quasi-Dynamic Approach in Structural Disorder Analysis: An Ion-Beam-Irradiated Silica. J. Phys. Chem. C 123 (2019) 29324-29330 http://dx.doi.org/10.1021/acs.jpcc.9b08895

[40] J. Tauc, Amorphous and Liquid Semiconductors; Plenum: New York, NY, USA, 1974.

[41] N.F. Mott, E.A. Davis, Electronic Processes in Non-Crystalline Materials; Oxford University Press: Oxford, UK, 1979.

[42] B. Dorner, H. Schober, A. Wonhas, M. Schmitt, D. Strauch. The phonon dispersion in 6H-SiC investigated by inelastic neutron scattering. Eur. Phys. J. B 5 (1998) 839–846. https://doi.org/10.1007/S100510050510